\documentclass[submission,copyright,creativecommons]{eptcs}
\providecommand{\event}{ML 2016} 

\usepackage{times}
\SetMathAlphabet{\mathtt}{normal}{OT1}{pcr}{m}{n}
\SetMathAlphabet{\mathtt}{bold}{OT1}{pcr}{bx}{n}

\usepackage{amsmath}
\usepackage{amssymb} 

\newcommand{\CUT}[1]{}

\newcommand{\secref}[1]{Section~\ref{#1}}

\newcommand{\figref}[1]{Figure~\ref{#1}}

\newcommand{\eg}{{\em e.g.}}

\newcommand{\ie}{{\em i.e.}}
\newcommand{\etc}{{\em etc.\/}}

\newcommand{\etal}{{\em et al.\/}}

\newcommand{\Cplusplus}{\mbox{C\hspace{-.05em}\raisebox{.4ex}{\tiny\bf ++}}}

\newcount\timeHH
\newcount\timeMM
\timeHH=\time
\divide\timeHH by 60
\timeMM=\time
\multiply\count255 by -60 \advance\timeMM by \count255
\newcommand{\timestamp}{%
  \today{} ---
  \ifnum\timeHH<10 0\fi\number\timeHH\,:\,\ifnum\timeMM<10 0\fi\number\timeMM}

\usepackage{graphicx}
\usepackage{listings}
\usepackage{dblfloatfix}
\usepackage[usenames,dvipsnames]{xcolor}
\usepackage{hyperref}
\usepackage{csquotes}
\usepackage{amsmath}
\usepackage{amsthm}
\usepackage{float}
\usepackage{xcolor}
\usepackage{textcomp}
\usepackage{wasysym}
\usepackage{pdfpages}
\usepackage{inconsolata}
\usepackage{graphicx}
\usepackage{amssymb}
\usepackage{setspace}
\usepackage{subcaption}
\usepackage{multicol}

\usepackage{underscore}           

\lstdefinelanguage{SML}{%
  morekeywords={%
    abstype, and, andalso, as, case, datatype, do, else, end, eqtype, exception,%
    fn, fun, functor, handle, if, in, include, infix, infixr, let, local, nonfix,%
    of, op, open, orelse, raise, rec, sharing, sig, signature, struct, structure,%
    then, type, val, where, while, with, withtype,%
  },%
  sensitive,%
  morecomment=[n]{(*}{*)},%
  morestring=[d]",%
}[keywords,comments,strings]%

\lstdefinelanguage
   [x64]{Assembler}     
   [x86masm]{Assembler} 
   {morekeywords={CDQE,CQO,CMPSQ,CMPXCHG16B,JRCXZ,LODSQ,MOVSXD, %
                  POPFQ,PUSHFQ,SCASQ,STOSQ,IRETQ,RDTSCP,SWAPGS, %
                  rax,rdx,rcx,rbx,rsi,rdi,rsp,rbp, %
                  r8,r8d,r8w,r8b,r9,r9d,r9w,r9b,
                  r10, r11, r12, r13, r14, r15,
                  movq, subq, addq, movabsq, retq}} 

\newcommand{\ghPurple}{\color[RGB]{167,29,94}}

\newcommand{\ghTeal}{\color[RGB]{1,134,179}}

\makeatletter
\lstdefinelanguage{llvm}{
  morecomment = [l]{;},
  morestring=[b]",
  sensitive = true,
  identifierstyle=\ttfamily,
  alsoletter={\%-@_*0123456789},
  classoffset=1, keywordstyle=\ttfamily\bfseries\ghPurple,
  morekeywords={
    define, declare, jwa, call, getelementptr, extractvalue, naked,
    if, goto, phi, compare, eq, add, mul, sub, return, undef
  },
  classoffset=2, keywordstyle=\ttfamily\bfseries\ghTeal,
  morekeywords={
    i64, i64*, void, i32, label
  },
  classoffset=3, keywordstyle=\ttfamily\bfseries\color{black},
  morekeywords={
    @foo, @invoke-gc, @factorial, @enterRTS, @genLabel, allocPair
  }
}
\makeatother

\title{Compiling with Continuations and LLVM}
\author{Kavon Farvardin \quad John Reppy
\institute{University of Chicago\\
Illinois, USA}
\email{\{kavon,jhr\}@cs.uchicago.edu}
}
\def\titlerunning{Compiling with Continuations and LLVM}
\def\authorrunning{K.\ Farvardin \& J.\ Reppy}

\begin{document}
\maketitle

\begin{abstract}
LLVM is an infrastructure for code generation and low-level optimizations, which
has been gaining popularity as a backend for both research and industrial compilers,
including many compilers for functional languages.
While LLVM provides a relatively easy path to high-quality native code, its design is
based on a traditional runtime model which is not well suited to alternative
compilation strategies used in high-level language compilers, such as the use of
heap-allocated continuation closures.

This paper describes a new LLVM-based backend that supports heap-allocated continuation
closures, which enables constant-time \texttt{callcc} and very-lightweight multithreading.
The backend has been implemented in the Parallel ML compiler, which is part of the Manticore
system, but the results should be useful for other compilers, such as Standard ML of
New Jersey, that use heap-allocated continuation closures.
\end{abstract}

%

\section{Introduction}

Maintaining a native code generator that targets multiple architectures is a hassle
for compiler writers that requires expert knowledge of each new processor's quirks.
Some functional-language compilers avoid this problem by targeting C as a
``portable assembly language''~\cite{sml2c,cmm}, but this approach has significant
drawbacks in both compile-time and runtime performance.
More recently, LLVM~\cite{lattner:masters-thesis,llvm-compilation-framework}, which
provides a low-level SSA-based representation
with many available optimization passes,
has emerged as a popular backend for compiler writers.
Although designed with imperative and object-oriented languages as its expected
clients, LLVM has been used to build backends for functional languages, such as
Standard ML~\cite{llvm:mlton}, SML\#~\cite{llvm:smlsharp}, Haskell~\cite{llvm:ghc}, and Erlang~\cite{llvm:erlang}.
While LLVM addresses the problem of maintaining native code generators and is a better
portable assembly language than C code, it still suffers from a bias toward C runtime
conventions, which makes it a less than ideal target for a functional-language
compiler.
Functional language implementations often use specialized register and calling
conventions,\footnote{
  For example, many implementations dedicate a specific register as an allocation
  pointer to support efficient open-coding of heap allocation.
} and require guaranteed tail-call optimization, mechanisms to communicate with
the garbage collector, and efficient support for features like first-class continuations.

In this paper, we present our approach to solving the problems of using LLVM as a backend
for functional-language implementations.
In particular, we show how to use LLVM to support the heap-allocated first-class continuation runtime
model~\cite{appel:cps-book} used by the SML/NJ system and by the Manticore
system~\cite{implicit-threading-in-manticore:jfp}.
We have integrated our approach into the \emph{Parallel ML} (PML) compiler that is part
of the Manticore project~\cite{implicit-threading-in-manticore:jfp}.
Initial observations suggest that the LLVM backend produces more efficient code
relative to the previous MLRisc~\cite{mlrisc} backend.

The LLVM backends for the Glasgow Haskell Compiler (GHC)~\cite{llvm:ghc}
and Erlang (ErLLVM)~\cite{llvm:erlang} use special language-specific
calling conventions added to LLVM that support tail-call optimization (TCO).
The LLVM backend for the MLton SML compiler uses trampolining to avoid the issues with TCO~\cite{llvm:mlton}.
As far as we know, no one has successfully implemented heap-allocated first-class continuations with LLVM.
In the remainder of the paper, we describe a new calling convention that we have added
to LLVM, how we handle the GC interface, and how we support capturing continuations to
support preemption.

%


\section{Background}

In this section, we provide an overview of Manticore's PML compiler, along with the high-level overview of how the LLVM backend was
integrated with the compiler.

\subsection{The PML Compiler}

The Parallel ML compiler (\textbf{pmlc}) that lies at the heart of the Manticore project
is implemented by a sequence of translations between intermediate languages (IRs)~\cite{manticore-ml07}.
There are six distinct IRs in the compiler:
\begin{enumerate}
  \item Parse tree --- the product of the parser.
  \item AST --- an explicitly-typed abstract-syntax tree representation.
  \item BOM --- a direct-style normalized $\lambda$-calculus.
  \item CPS --- a continuation-passing-style $\lambda$-calculus.
  \item CFG --- a first-order control-flow-graph representation.
  \item MLTree --- the expression tree representation used by the
    MLRISC code generation framework~\cite{mlrisc}.
\end{enumerate}%

\begin{figure}
  \begin{center}
    \includegraphics[width=0.67\textwidth]{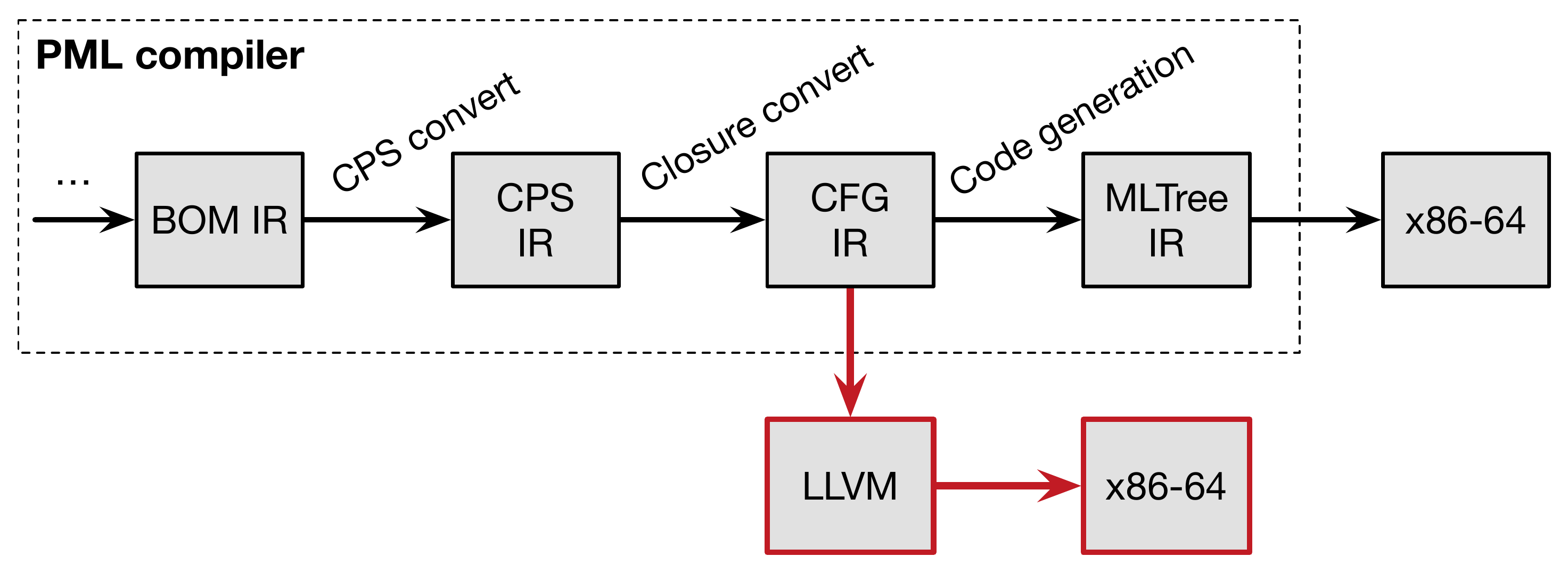}
  \end{center}
  \caption{The middle and backends of the PML compiler.  The new code-generation path is
    highlighted in red.}
  \label{fig:pipeline}
\end{figure}

We support the parallelism and concurrency features of PML by transformations
on the AST and BOM representations that introduce explicit continuation
binders~\cite{sched-framework-for-parallel,implicit-threading-in-manticore:jfp,data-only-flattening,parallel-cml}.
The translation from BOM to CPS uses the Danvy-Filinski CPS transformation~\cite{representing-control:cps-study}
and we perform several kinds of optimizations on the CPS IR~\cite{arity-raising,effective-ho-optimizations}.
The conversion from CPS to CFG uses a flat, safe-for-space, closure representation~\cite{cardelli1983}.
All closures, including those for return continuations, are immutable and allocated on the heap.
The CFG IR produced through closure conversion is still in continuation-passing style, but functions are no longer nested.
Finally, we generate an MLTree representation from the CFG and use the MLRISC code generation framework
to handle instruction selection and register allocation.
In this paper, we describe our experience with replacing the MLRISC code generation framework
by LLVM~\cite{llvm-compilation-framework} as illustrated in \figref{fig:pipeline}.

\begin{figure}
  \begin{center}
\begin{lstlisting}[language=SML,xleftmargin=5em,multicols=2]
datatype ty
  = Raw of machine_ty
  | Tuple of ty list
  | Addr of ty
  | Fun of ...
  ...
type var = id * ty
datatype func = FUNC of {
  entry : convention,
  start : block,
  body : block list
}
and block = BLK of {
  name : label,
  args : var list,
  body : exp list,
  exit : transfer
}
and transfer
  = ApplyFun of ...
  | ThrowCont of ...
  | If of ...
  | HeapCheck of ...
  ...
\end{lstlisting}
  \end{center}
  \caption{
    Simplified SML type definitions for the CFG IR.
  }
  \label{fig:cfgir}
\end{figure}

\subsection{The CFG Representation}
The CFG IR is a first-order, machine-like representation whose primary construct is the basic block (Figure \ref{fig:cfgir}).
Functions consist of a calling convention, a start block, and a list of body blocks.
Each block ends in transfer to another block within the function or a tail-call to either a function or continuation.
Throws to the continuation of a local control-flow divergence (\ie, a local join point) are identified during closure
conversion and are translated as a transfer to a block within the enclosing function.

In preparation for generating native code that uses bump allocation in the heap, we insert heap-limit tests into the
CFG representation.
We use GC tests as safe-points where preemptive thread switching can occur~\cite{reppy:signals}.
Thus, we must guarantee that every loop (even ones that do not allocate) includes at least one heap limit test.
To ensure every loop has a limit test, we use a bounded algorithm that minimizes the feedback vertex set of
the control-flow graph of the program.
The limit tests introduce continuation captures that are not present in the original program,
which can be used to implement preemptive scheduling (Section \ref{sec:cont}).

\subsection{Interfacing with LLVM}
There are several ways to generate code for LLVM.
One can generate code directly using either the native \Cplusplus{} APIs, which gives one full access to all of the LLVM
features, or the more restricted C APIs, which do not support some newer features, such as \texttt{musttail} calls.
Lastly, one can generate LLVM assembly code and use the \textbf{llc} command-line tool to generate assembly code.
Since there are no SML bindings for LLVM in the SML/NJ system that we use to develop
our code, we built a library for generating LLVM assembly code that is compatible with LLVM 3.8 or newer.

%

\section{Translating to LLVM IR}

In this section, we provide complete details of the translation from the CFG IR to LLVM IR,
while touching on aspects of our runtime system and LLVM's code generation.

%

\subsection{Conversion to SSA}
The LLVM IR is a static-single-assignment (SSA) representation, but it is possible
to generate non-SSA code and allow LLVM to do the SSA conversion for you.
The approach, which is used by the MLton compiler~\cite{llvm:mlton}, is
to stack allocate variables and emit \texttt{load/store} instructions whenever accessing them.
The \texttt{mem2reg} pass in LLVM will then promote these values to virtual register
accesses, using good heuristics to insert $\phi$-nodes in the program as-needed~\cite{llvm:passes}.

The CFG IR is already in a static single-assignment form, but data-flow joins within
a function do not use an explicit $\phi$-node to merge their values as in a typical
SSA representation.
Instead, the CFG IR takes a functional approach by assigning parameters to each block,
with the predecessors of a block making up the set of possible values for each
parameter~\cite{appel:ssa}.
Our library to generate LLVM follows suit by using parameterized blocks and requiring
the user to note incoming block transfers.
With this information, it is straightforward to emit $\phi$-nodes for each block.
While this produces a $\phi$-node per parameter in each block, the \texttt{instcombine}
optimization pass in LLVM eliminates redundant nodes that simply rename a value.

\subsection{Type Correspondences}

The main base types in the CFG IR, \ie{}, integers, floats, and addresses, have an obvious corresponding type in LLVM.
However, there are two separate difficulties when representing tuples and functions in LLVM.

Ideally, a tuple of values in the CFG IR corresponds to a pointer to a struct in LLVM, providing richer type information for optimization.
But, our representation of tuples in memory uses a field alignment convention that deviates from those used in languages such as C.
While an LLVM module can specify a \texttt{datalayout} string that describes such alignments, this string is only used to inform LLVM's optimizer of what the \emph{code generator} will do.
It is currently not possible to override the code generator's memory alignment assumptions for struct fields.
Thus, all tuples are represented as a pointer to an 8-byte integer so that all fields are 8-byte aligned; values are casted as needed for loads and stores involving field pointers.

There are two factors that constrain the type of a CFG function in LLVM.
The first is that our runtime system expects certain register conventions for function arguments, so we must order them specially (Section \ref{sec:gc}).
The second is that tail calls emitted by our compiler \emph{must} remain as tail calls, so we use the \texttt{musttail} call marker (Section \ref{sec:tco}).
A \texttt{musttail} call requires that the types of the caller and callee match, modulo the type pointed to by a pointer~\cite{llvm:musttailProposal}.

Since all invocations of CFG functions and continuations are tail calls, and Manticore's runtime system uses more than one register convention to pass arguments to functions, we must carefully match up the LLVM types used by caller and callee.
Every CFG function is given a new type in LLVM by padding the parameter list with 64-bit integers, so that every general-purpose register that could ever be used for passing arguments will be consumed by that type.
Functions of this type allow us to skip function casts and implement multiple register conventions without needing more than one LLVM calling convention, since we can reorder parameters so they end up in the right machine register.
We pass \texttt{undef} values for arguments not used by the callee, and these \texttt{undef} values disappear when LLVM generates machine code.

%

\subsection{Proper Tail-Call Optimization}
\label{sec:tco}

Most functional languages use tail recursion to express loops and, thus, require
that tail calls do not grow the stack.
In limited circumstances, LLVM can meet this requirement, but even when LLVM is able
to apply TCO to a call, there is extra stack-management overhead on function entry and
exit~\cite{llvm:tailUnreachable,llvm:musttailUnreachable,llvm:inefficienttail}.

This extra overhead is bad enough for implementations that rely on traditional
stacks, but it is prohibitively expensive for implementations that use CPS and
heap-allocated continuations~\cite{appel:cps-book}.

A standard technique to avoid this overhead is to merge mutually tail-recursive
functions into a single LLVM function and then use internal jumps instead of
tail calls.
Unfortunately, this approach does not work for tail calls to unknown functions
and incurs substantial compile-time cost.
The MLton SML compiler partitions the generated code into multiple large functions
(called \emph{chunks}) and uses a trampoline to transfer control between
chunks~\cite{llvm:mlton}.
This approach solves the compile-time issue, but places extra strain on the
register allocator because of the nature of the control-flow graphs in the chunks.

Our solution to this problem is to add a new calling convention, which we
call Jump-With-Arguments (JWA), to LLVM.
This calling convention has the property that it uses all of the available
hardware registers and that no registers are preserved by either the caller
or callee.
Furthermore, the argument registers are exactly the live registers on return
(\ie{}, the call and return have identical register use conventions).\footnote{
  These properties are why we need to create a new convention, instead of using
  an existing convention.
}
The JWA convention is not specific to any architecture or version of LLVM, though
we have currently only implemented it for x86-64 in a fork of LLVM to support
the PML compiler.

\begin{figure}
\centering
\includegraphics[width=0.45\textwidth]{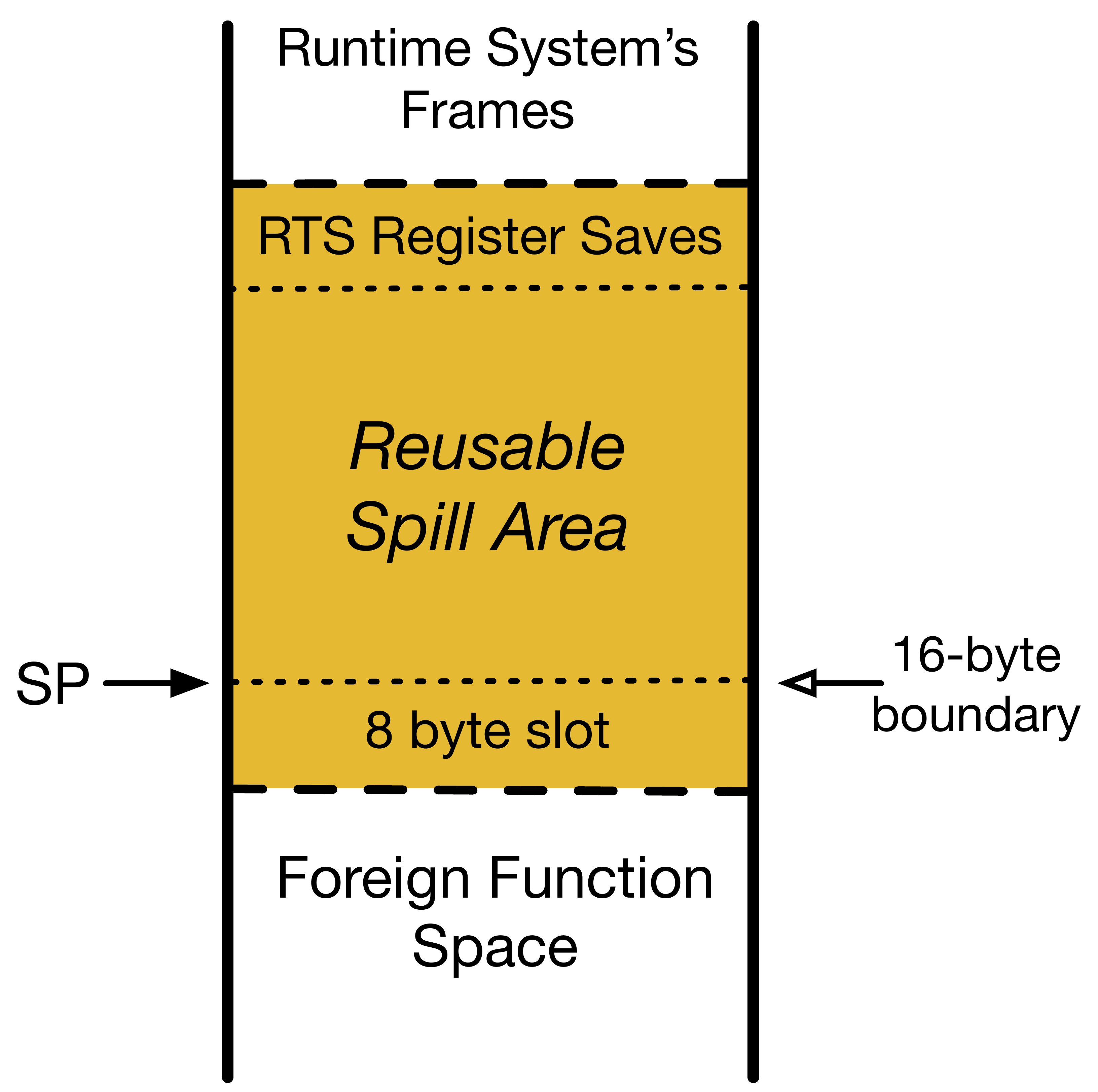}
\caption{
    The shaded frame is setup by runtime system and is shared by all PML functions.
}
\label{fig:nakedframe}
\end{figure}

Second, we mark every function with the \lstinline!naked! attribute,
which tells the code generator to completely skip the emission of a
function prologue and epilogue.
This attribute \emph{must} be used with care; it was originally
designed to support functions consisting
entirely of inline assembly that manages the stack explicitly (\eg{},
interrupt service routines).
Using the \lstinline!naked! attribute means that the generated code
is responsible for ensuring that there is sufficient stack space
for register spills and any callee-saved registers are preserved.

We use an assembly language shim for switching between the runtime
system code (written in C) and the code generated by the PML compiler.
This shim code allocates a frame (Figure \ref{fig:nakedframe}) that is
large enough to handle the maximum number of register spills.
This technique is borrowed from the SML/NJ system; the spill limit is
enforced by the compiler limiting the number of live variables at any control point and over-provisioning the spill area.\footnote{
    As there is only one spill area per hardware thread, allotting a few kilobytes is no problem.
}
In short sequences, LLVM's code generator may introduce one or two additional spills, but by limiting the number of values early on, we effectively bound the number of register spills.
LLVM's code generator will assign any register spills to frame locations
starting from the bottom of the frame, offset from the stack pointer, so
the shim preserves all C callee-saved registers at the top of the frame.

%

\subsection{Allocation and Garbage Collection}
\label{sec:gc}

Our JWA calling convention maps function arguments to hardware registers based on the
position of the argument.
By using standard positions for special runtime registers, we can effectively pin them
to hardware registers (\eg{}, we always pass the allocation pointer as the first
argument).
Object allocation then defines new instances of the allocation pointer, which thread
the current state of the pointer through the code (recall that LLVM code is in SSA form).


One of the advantages of CPS with heap-allocated continuations is that the interface
to garbage collection is very simple.
The runtime does not need to scan a stack (since there is no stack) or understand
any other properties of the code generator.
We did not have to make any modifications to our existing collector to support our LLVM backend.

The compiler is responsible for generating code to check for heap exhaustion and
code to invoke the GC when necessary.
In \figref{fig:gc}, we list simplified LLVM code for the heap-limit check
(Lines~4--5) and GC invocation.
\begin{figure}[t]
  \begin{center}
    \parbox{0.8\textwidth}{
\begin{lstlisting}[language=llvm]
declare jwa {i64*, i64*} @invoke-gc(i64*, i64*)

define jwa void @foo(i64 allocPtr_0, ... ) naked {
...
  if enoughSpace, label continue, label doGC
    
doGC:
  roots_0 = allocPtr_0
  ; ... save live vals in roots_0 ...
  allocPtr_1 = getelementptr allocPtr_0, 5 ; bump
  retV = call jwa {i64*, i64*} 
           @invoke-gc(allocPtr_1, roots_0)
  allocPtr_2 = extractvalue retV, 0
  roots_1 = extractvalue retV, 1
  ; ... restore live vals ...
  goto label continue
    
continue:
  allocPtr_3 = phi i64* [ allocPtr_0, allocPtr_2 ]
  liveVal_1 = phi i64* [ ... ]
...
\end{lstlisting}}
  \end{center}%
  \caption{
    An example of a compiler generated safe point for garbage collection.
  }
  \label{fig:gc}
\end{figure}%
To invoke the GC, we first save the live variables into a new heap object called \texttt{roots} using bump-allocation (Lines~8--10)
and then do a non-tail JWA call to \lstinline!@invoke-gc!.
When this function returns, we restore the allocation pointer and live variables  (Lines~13--15).

We use a non-tail call to \lstinline!@invoke-gc! for reasons described in below in \secref{sec:cont}.
We ensure that LLVM does not try to preserve values across the \lstinline!@invoke-gc!
call by taking advantage of the rules about aliasing.
Once the pointer reaching all live values is passed to
\lstinline!@invoke-gc!, which is an external function not visible to LLVM,
LLVM must assume that all values may have changed and must use the updated versions
from the pointer returned.







%

\subsection{Preemption and Multithreading}
\label{sec:cont}

The main motivation for supporting heap-allocated first-class
continuations is to enable the efficient implementation of
the concurrency mechanisms in the Manticore runtime
system~\cite{sched-framework-for-parallel,manticore:stm}.
While the mechanisms described in \secref{sec:tco} are sufficient
to support the explicit management of continuations, preemptive
scheduling requires capturing continuations that are not explicit
in the intermediate code.
We use the technique developed for supporting asynchronous signals
in SML/NJ~\cite{reppy:signals}, which limits preemption to
\emph{safe points} where all live values are known.
Specifically, those places in the code where we perform a heap-limit check
serve as safe points.\footnote{
  The code generator ensures that even non-allocating loops contain
  a heap-limit check.
}
We store the heap limit pointer in memory, which means that we can
set it to zero when we need to force a heap-limit check to fail.
The runtime system then constructs a continuation closure, which is
passed to the preemption handler where it can be put on scheduling
queue \etc{}

This mechanism introduces an additional challenge for our LLVM
backend, because the implicit continuations that are captured
during preemption do not correspond to LLVM functions and
are invoked from unknown locations.
For example, consider the heap-limit test in \figref{fig:gc}.
If it is invoked to force a preemption, then a continuation will
be created by the runtime system that has Line~13
as its code address (\ie{}, the return address of the call to
\lstinline!@invoke-gc!).
Since Line~13 is not a function entry, there is no way in LLVM to
specify a calling convention.

We are saved, however, by the fact that we can specify our own return convention
for structs in LLVM.
Normal conventions return a struct using a mix of registers or stack space that does not match up with the way arguments are passed to functions.
We setup our JWA convention so that struct field $i$
is returned in the same register that argument $i$ would be passed in during a call.
This way, return addresses generated through a non-tail JWA call can be jumped to safely using a standard JWA tail call (which is how we throw to a continuation!).

\begin{figure}[t]
  \centering
  \begin{subfigure}[b]{0.8\textwidth}
\begin{lstlisting}[language=llvm]
define jwa void @foo( ... ) naked {
...
preempted:
  env = ; ... save live vars ...
  closPtr = allocPair (undef, env)
  ret = call jwa {i64*, i64*} 
            @genLabel(closPtr, @enterRTS)
  arg1 = extractvalue ret, 0
  arg2 = extractvalue ret, 1
...
}
\end{lstlisting}
      \caption{
        A non-tail call to produce a ``block'' label.
      }
      \label{fig:callcc:preemption}
  \end{subfigure}
  \\
  \begin{subfigure}[b]{0.8\textwidth}
\begin{lstlisting}[language=llvm]
; call convention:
; rsi = closPtr, r11 = @enterRTS
genLabel:
  pop rax       ; pop ret addr
  mov rax,(rsi) ; finish closure
  jmp r11
    
\end{lstlisting}
      \caption{
        Assembly shim to pick up the label from the stack.
      }
      \label{fig:callcc:genlabel}
  \end{subfigure}%
  \caption{
  An example of generating a first-class label in LLVM.
  }
  \label{fig:callcc}
\end{figure}%

To help illustrate this point, consider the example in \figref{fig:callcc}.
On lines 4--5, we save all live values into a new heap-allocated closure,
but the code pointer is left uninitialized.
This incomplete closure along with the function we intend to invoke are
passed in a non-tail call to \lstinline!@genLabel!.
The key point is that this call will push a return address on the stack
that resumes on line 8, which resumes execution by using the values returned.
We ensure that there are no values live across the non-tail call, so the
stack frame of the caller, \lstinline!@foo!, can be safely reused.
Then, the assembly shim \lstinline!@genLabel! pops the return address,
which has a calling convention that is identical to a call, and places
it into the closure before invoking \lstinline!@enterRTS!.

%

\subsection{LLVM Optimizations}
\label{sec:opt}

One of the benefits of using LLVM is that it provides a rich set of optimization passes and opportunities to tune the output of its code generator.
A particularly beneficial optimization is to guide LLVM's basic block placement for heap limit tests, since these tests normally fail (\ie{}, the heap is not exhausted). 
It is important to keep basic blocks that handle overflow away from the hot path of the function to reduce instruction cache pressure.
We use the built-in \lstinline!@llvm.expect! intrinsic along with the \lstinline!lowerexpect! pass to add branch probabilities to these tests, which guides the code generator when performing block layout.

We also hand-crafted two sequences of LLVM optimization passes, using trial-and-error guided by intuition,\footnote{
  Ideally, we would automate this process~\cite{autotune-passes}.
} to simplify the LLVM program output by the PML compiler.
Crafting a custom pass sequence is recommended for compiler frontends that target LLVM, as the default ``-O$x$'' passes are tuned for a C/C++ frontend~\cite{llvm:advice}.
Our ``Basic'' optimization will only shrink the size of the program, and consist of a combination of the following passes:
\lstinline!simplifycfg!,
\lstinline!instcombine!,
\lstinline!reassociate!,
\lstinline!constprop!,
\lstinline!early-cse!,
\lstinline!gvn!, and
\lstinline!dce!.
``Extra'' optimization adds passes to the ``Basic'' suite that can specifically optimize memory operations, as they are very common:
\lstinline!sink!,
\lstinline!mldst-motion!, and
\lstinline!slp-vectorizer!.

%

\section{Evaluation}
\label{sec:eval}

\begin{figure}
\centering
\includegraphics[width=0.98\textwidth]{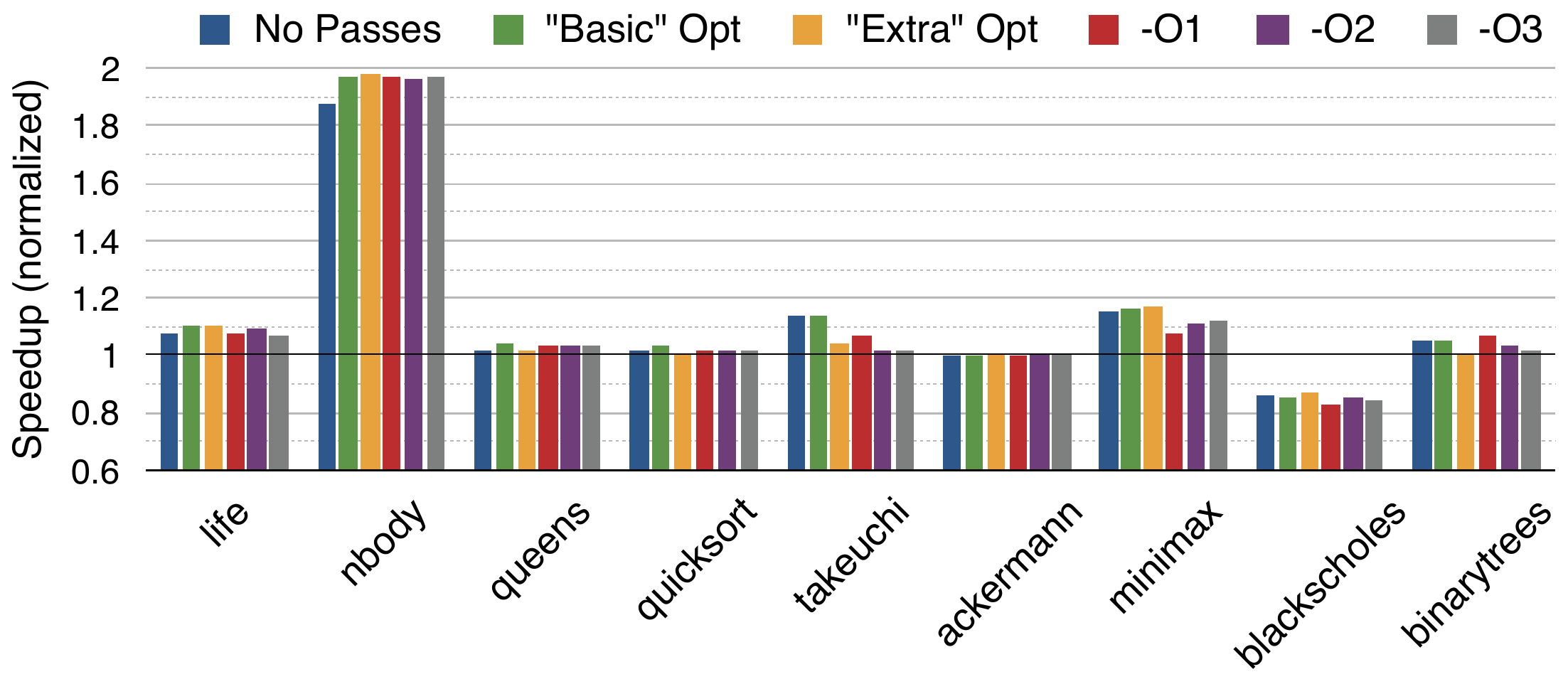}
\caption{
    Execution time speedups when compiled with LLVM, normalized to MLRisc. Each bar represents a different set of additional optimizations applied when compiling with LLVM.
}
\label{fig:eval}
\end{figure}

We measured the difference in application performance between our two backends to get a sense
of how well LLVM can handle the unusual code we generate. 
\figref{fig:eval} summarizes our experiment conducted on a workstation equipped with two
E5-2687W processors and 64GB of memory. 
Speedups reported are relative to the MLRisc backend and were computed using the average of 50 trials.
Error bars are omitted as the standard error was less than 1.5\%.
Each benchmark was compiled with different sets of additional LLVM optimization passes applied before generating assembly (\secref{sec:opt}).
Optimizations of the form ``-O$x$'' indicate one of the default optimization levels built into LLVM.

The significantly better performance by LLVM over MLRisc on the nbody benchmark is owed to poor
use of floating-point registers.
We have recently identified a mistake in the way that the MLRisc does register allocation for the x86-64,
which results in significantly more register shuffling.
We have not yet had the opportunity to fix MLRisc, so we do not know how big of a difference this fix
will make in the results.

On the other hand, we have not found a reason for the notably worse performance when using LLVM for the parallel blackscholes benchmark.
When testing this benchmark on a 2013 Macbook Pro, the LLVM-generated version matches or outperforms the MLRisc version.
Thus, the blackscholes performance may be sensitive to the particular CPU(s) on the machine.

While the takeuchi benchmark only tests the overhead of recursion, its performance takes a hit with aggressive optimization in LLVM because of the \texttt{slp-vectorizer} pass. 
After the pass is applied, the hot path is smaller because of the use of vector instructions to initialize a closure, but the execution cost of such instructions outweighed the size benefit.
Overall, LLVM seems to produce better code than MLRisc does, with some programs making significant gains.

%

\section{Related Work}
In general, preemption and explicit stack management are treated similarly in the GHC and PML compilers.
GHC uses a whole-program CPS transformation to make the call stack explicit in its first-order intermediate representation, regardless of the backend being used.
But because of the lack of first-class labels in LLVM, GHC is forced to perform an additional splitting transformation (\ie{}, ``proc-point splitting'') at every return point, such as safe points~\cite{procpoint}.
A solution similar to the technique described in \secref{sec:cont} can remove the need for a splitting transform when targeting LLVM.

Dolan \etal\ proposed the \texttt{SWAPSTACK} mechanism for LLVM to enable lightweight context switching~\cite{swapstack}.
To capture stack-allocated one-shot continuations in LLVM, a mechanism like \texttt{SWAPSTACK} can be used in conjunction with runtime system support.
While our focus is on fully general first-class continuations via immutable heap-allocated frames, our mechanism for \lstinline!@genLabel! (\figref{fig:callcc}) is similar in spirit to \texttt{SWAPSTACK}.
A major difference is that we leverage liveness properties of a program in continuation-passing style to correctly implement \lstinline!@genLabel!.
In addition, \emph{explicit} continuation captures in the original program do not need our mechanism at all, thus avoiding the runtime system.

\section{Conclusion and Future Work}
We have outlined how to extend LLVM to support the heap-allocated first-class
continuation runtime model.
We are in the process of replacing the MLRisc backend with LLVM using
the approach described in this paper.
Initial observations suggest that this new LLVM backend produces smaller and
more efficient code.
We are also hoping to apply these techniques to the SML/NJ system, and integrate a generalized form of this work into LLVM in the future.

\section*{Acknowledgments}
This material is based upon work supported, in part, by the National Science Foundation
under Grant CCF-1010568. The views and conclusions contained herein are those
of the authors and should not be interpreted as necessarily representing the
official policies or endorsements, either expressed or implied, of these
organizations or the U.S.\ Government.

\bibliographystyle{eptcs}
\bibliography{strings-short,refs,manticore}

\end{document}